\documentstyle[12pt]{article}
\def\kmu{K_L\to\mu\bar{\mu}}
\def\kpmu{K^+\to\pi^+\mu^+\mu^-}
\def\kmn{K^+\to\pi^0\mu^+\nu}

\def\kmu2g{K^+\to\mu^+\nu\gamma}
\newcommand{\nn}{\nonumber}
\newcommand{\bd}{\begin{document}}
\newcommand{\ed}{\end{document}}
\newcommand{\bc}{\begin{center}} 
\newcommand{\ec}{\end{center}}
\newcommand{\be}{\begin{eqnarray}}
\newcommand{\ee}{\end{eqnarray}}
\newcommand{\eqn}{\global\def\theequation}
\newcommand{\sw}{sin^2 \theta_W}
\newcommand{\fbd}{f_B}
\renewcommand{\thefootnote}{\alph{footnote}}
\newcommand{\se}{\section}
\newcommand{\sse}{\subsection}
\newcommand{\bi}{\bibitem}
\begin{document}
\tolerance=10000
\baselineskip=7mm
\begin{titlepage}  

 \vskip 0.5in   
 \null
\begin{center}
 \vspace{.15in}
{\LARGE {\bf T Violating Muon Polarization in $K^+\to \mu^+\nu\gamma$}
}\\
\vspace{1.0cm}
  \par
 \vskip 2.5em
 {\large
  \begin{tabular}[t]{c}
{\bf C.~H.~Chen$^a$, C.~Q.~Geng$^b$ and C.~C.~Lih$^b$}
\\
\\
       {\sl ${}^a$ Institute of Physics, Academia Sinica}\\
       {\sl  $\ $ Taipei, Taiwan,  Republic of China }\\
\\
and\\
\\ {\sl ${}^b$ Department of Physics, National Tsing Hua University} 
\\  {\sl  $\ $ Hsinchu, Taiwan, Republic of China }\\
   \end{tabular}}
 \par \vskip 5.0em
 {\Large\bf Abstract}
\end{center}

We study the T violating transverse muon polarization in the decay of
$K^+\to \mu^+\nu\gamma$ due to CP violation in theories beyond the 
standard model. We find that the polarization asymmetry could be large 
in some CP violation models and it may be detectable at the ongoing KEK 
experiment of E246 as well as the proposed BNL experiment.

\end{titlepage}

\se{Introduction}
$\ \ \ $

In the framework of local quantum field theories, with Lorentz invariance 
and the usual spin-statistics connection, time-reversal (T) violation
implies CP violation (and vice versa), because of the CPT invariance of 
such theories. Experimentally, only CP violation has been observed so 
far and this only in the neutral kaon system.
But the origin of this violation remains unclear.  
In the standard model, CP Violation arises from a unique
physical phase in the Cabibbo-Kobayashi-Maskawa (CKM) quark mixing 
matrix \cite{ckm}. To ensure that this phase is indeed the source of CP 
violation or T violation, one needs to look for new processes,
especially that outside $K^0$ system. It would be particularly interesting if
the time reversal symmetry is directly violated in these processes,
rather than inferring it as a consequence of CPT invariance.

Within kaon physics, the most interesting search of T violation
would be to look for the component of the muon polarization 
normal to the decay plane, called transverse muon polarization ($P_T$), 
in the charged kaon decays such as $\kmn$ ($K^+_{\mu 3}$) \cite{Sakurai}, 
$\kmu2g$ ($K^+_{\mu 2\gamma}$) \cite{Bryman} and $\kpmu$ \cite{ANBG}. 
The polarization in $K^+_{\mu 3}$ as well as that in $K^+_{\mu 2\gamma}$
will be measured to a high accuracy at the ongoing KEK E246
experiment \cite{kuno_p} and at a recently proposed BNL 
experiment \cite{BNL}, while for $\kpmu$ that would be done 
in a future kaon factory \cite{kuno-k}. 

In this paper, we concentrate on the radiative $K^+_{\mu 2}$ decay.
We will first give a general analysis on all components of the muon 
polarization and then present our estimations on the transverse 
component in various CP violation theories beyond the standard model.
The transverse muon polarization in the decay of $\kmu2g$
is related to the T odd triple correlation, $i.e.$,
\be
P_T &\propto &
\vec{s}_{\mu}\cdot (\vec{p}_{\mu}\times \vec{p}_{\gamma})\,,
\label{eqn:1}
\ee
where $\vec{s}_{\mu}$ is the muon spin vector and $\vec{p}_{i}\,,\ 
(i=\mu$ and $\gamma$) the momenta of the muon and photon
in the rest frame of $K^+$, respectively.
It is expected \cite{geng-kek} that the CKM phase does not induce
the polarization in Eq. (\ref{eqn:1}).
Therefore measuring the polarization could be a signature of physics 
beyond the standard model.
There are many different sources that might give rise to 
the polarization. The most exciting ones are the weak CP violation 
from some kinds of non-standard CP violation models.
However, the electromagnetic interaction among the final state particle
can also make a contribution, which is usually less interesting
and it could even hide the signals from the weak CP violation.
We shall refer the final-state-interaction (FSI) contribution
as a theoretical background of that by the weak CP violation.
It has been estimated that 
$P_{T}\sim 10^{-3}$ and $10^{-6}$ 
in $K_{\mu3}^0\ (K^0\rightarrow\pi^-\mu^+\nu$) \cite{FSI1}
and $K_{\mu 3}^+\ (K^+\to \pi^0\mu^+\nu)$ \cite{FSI2}, due to the 
FSI effects at one and two-loop levels,
respectively. For $\kmu2g$, although there is only one charged final 
state particle like $K_{\mu 3}^+$, FSI arises at one-loop diagrams 
because of the existence of the photon in the final states \cite{Marciano,g1}
and it is found that $P^{FSI}_{T}(\kmu2g)\sim 10^{-3}$ in most of the 
decay allowed phase space \cite{gly}.  To distinguish the real
CP violating effects from the FSIs, one has to explore various 
possible models with the polarization being larger than $10^{-3}$.

The paper is organized as follows. In Sec. 2, we carry out a 
general analysis of the muon polarization in $K^+\to\mu^+\nu\gamma$.
In Sec. 3, we study the CP violating muon polarization effects in some 
extensions of the standard model. We give our concluding remarks 
in Sec.~4. 

\se{General Analysis for Muon Polarization}
\ \ \ 
 For a general investigation of the transverse muon polarization 
in $K^+\to \mu^+\nu\gamma$ decay including
some new CP violating sources, we first carry out 
the most general four-Fermion interactions given by
\be
{\cal L}&=&-{G_F \over \sqrt{2}}sin\theta_c \bar{s}\gamma^{\alpha}
(1-\gamma_5) u
\bar{\nu}\gamma_{\alpha}(1-\gamma_5)\mu+ G_S 
\bar{s}u \bar{\nu}(1+\gamma_5)\mu 
+G_P \bar{s}\gamma_5 u \bar{\nu}(1+\gamma_5)\mu \nonumber \\
&+&G_V \bar{s}\gamma^{\alpha}u
\bar{\nu}\gamma_{\alpha}(1-\gamma_5)\mu+G_A \bar{s}\gamma^{\alpha}\gamma_5 u
\bar{\nu}\gamma_{\alpha}(1-\gamma_5)\mu+h.c.,  
\label{eqn:fermi}
\ee
where $G_F$ is the Fermi constant, $\theta_c$ is the Cabibbo mixing angle, and
$G_S$, $G_P$, $G_V$, and $G_A$, arising from new physics,
denote scalar, pseudoscalar, vector, and axial vector interactions, 
respectively. 
 From the interactions in Eq. (\ref{eqn:fermi}),
we can write the amplitude of the decay $K^+_{\mu 2\gamma}$ 
in terms of ``inner bremsstrahlung'' (IB) and ``structure-dependent'' (SD) 
contributions, which can be written as \cite{Bryman,Bijnens}
\be
M=M_{IB}+M_{SD}, \nonumber 
\ee
with
\be
M_{IB}&=&i e{G_F \over \sqrt{2}} sin\theta_c f_K m_{\mu}
 \epsilon^*_{\alpha}K^{\alpha}, 
\nn
\\
M_{SD}&=&-i e{G_F \over \sqrt{2}} sin\theta_c \epsilon^*_{\mu}L_{\nu}H^{\mu\nu},
\label{eqn:sd}
\ee
where
\be
K^{\alpha}&=&\bar{u}(p_{\nu})(1+\gamma_5)\left ({p^{\alpha} \over p\cdot q}
-{2 p^{\alpha}_{\mu}+ \not\!q \gamma^{\alpha} \over 2 p_{\mu}\cdot q}
\right)v(p_{\mu},s_{\mu}),
\nonumber \\
L_{\nu}&=&\bar{u}(p_{\nu})\gamma_{\nu}(1-\gamma_5)v(p_{\mu},s_{\mu}), 
\nonumber\\ 
H^{\mu\nu}&=&{F_A \over m_K}\left(-g^{\mu\nu}p\cdot q+p^{\mu}q^{\nu}\right)+
i{F_V \over m_K}\epsilon^{\mu\nu\alpha\beta}q_{\alpha}p_{\beta}\,,
\ee 
$\epsilon_{\alpha}$ is the photon polarization vector,
$p$, $p_{\nu}$, $p_{\mu}$, and $q$ are the four-momenta of
$K^+$, $\nu$, $\mu^+$, and $\gamma$, respectively, 
$s_{\mu}$ is the polarization vector of the muon,
and $f_K$, $F_V$, 
and $F_A$ are the form factors given by 
\be
f_K
&=& f^0_K \left(1+\Delta_P +\Delta_A\right),
\nn\\
 F_A&=&
F^0_A (1+\Delta_A),
\nn\\
 F_V&=&
F^0_V (1-\Delta_V),
\label{eqn:fk} 
\ee
with
\be
\Delta_{(P,A,V)}&=&
{\sqrt{2}\over G_F sin\theta_c}
\left({G_Pm^2_K \over (m_s+m_u)m_{\mu}},G_A,G_V\right)\,.
\label{eqn:fkp}
\ee
Here $f^0_K$ is the kaon decay constant and
$F^0_{V(A)}$ the vector (axial-vector) form factor,
defined by
\be
<0|\bar{s}\gamma^{\mu}\gamma_5 u|K^+(p)> &=&-if^0_Kp^{\mu},
\nn
\\
\int dx e^{iqx}<0|T(J^{\mu}_{em}(x)\bar{s}\gamma^{\nu}\gamma_5 u(0))|K^+(p)>
&=&
-f^0_K\left ( g^{\mu\nu}+{p^{\mu}(p-q)^{\nu} \over p\cdot q}\right)
\nn
\\
&&+{F^0_A \over m_K}(g^{\mu \nu} p\cdot q-p^{\mu}q^{\nu}),
\nn
\\
\int dx e^{iqx}<0|T(J^{\mu}_{em}(x)\bar{s}\gamma^{\nu} u(0))|K^+(p)>
&=&
i{F^0_V \over m_K}\epsilon^{\mu\nu\alpha \beta}q_{\alpha}p_{\beta}, 
\ee
and the
hadronic matrix elements involving the scalar and pseudoscalar currents in
Eq. (\ref{eqn:fk}) are given by \cite{Kobayashi}
\be
<0|\bar{s}\gamma_5 u|K^+(p)>&=&if^0_K {m^2_K\over m_s+m_u},
\nn\\
\int dx e^{iqx}<0|T(J^{\mu}_{em}(x)\bar{s}\gamma_5 u(0))|K^+(p)>
&=&
f^0_K{m^2_K \over m_s+m_u}{p^{\mu} \over p\cdot q}, 
\nn\\
\int dx e^{iqx}<0|T(J^{\mu}_{em}(x)\bar{s}u(0))|K^+(p)>
&=&0\,,
\ee
where $J^{\mu}_{em}$ is the electromagnetic current. 
Numerically, one has $f^0_K=0.16\ GeV$ from the experiment and
$F^0_V=-0.095$ and $F^0_A=-0.043$ found in the chiral perturbation   
theory at one-loop level \cite{Bijnens}.

We use the standard techniques to calculate the probability of the process 
$K^+ \to\mu^+ \nu\gamma$ as a function of the 4-momenta of the particles 
and the polarization 4-vector $s_{\mu}$ of the muon.
We write the components of $s_{\mu}$
in terms of $\vec{\xi}$, a unit vector along the muon spin in its rest 
frame, as
\be
s_0={\vec{p}_{\mu}\cdot \vec{\xi} \over m_{\mu}},\ \ \vec{s}=\vec{\xi}+
{s_0 \over E_{\mu}+m_{\mu}}\vec{p}_{\mu}.
\ee  
In the rest frame of $K^+$, 
the partial decay width is found to be
\be
d\Gamma={1\over 2m_K}|M|^2 (2\pi)^4\delta(p-p_{\mu}-p_{\nu}-q)
{d\vec{q}\over (2\pi)^3 2E_{q}} {d\vec{p}_{\mu}\over (2\pi)^3 2E_{\mu}}
{d\vec{p}_{\nu}\over (2\pi)^3 2E_{\nu}},
\ee
with
\be
|M|^2=\rho_0(x,y)[1 + (P_L\vec{e}_L +P_N\vec{e}_N + P_T\vec{e}_T)\cdot 
\vec{\xi}\,]
\ee
where 
$\vec{e}_i\ (i=L,N,T)$ are the unit vectors along the 
longitudinal, normal and transverse  components of the
muon polarization, defined by
\be 
\vec{e}_L&=&{\vec{p}_{\mu} \over |\vec{p}_{\mu}|},
\nn\\ 
 \vec{e}_N&=&{\vec{p}_{\mu}\times (\vec{q}\times \vec{p}_{\mu})
 \over |\vec{p}_{\mu}\times (\vec{q}\times \vec{p}_{\mu})|},
\nn\\ 
 \vec{e}_T&=&{\vec{q}\times \vec{p}_{\mu}\over 
|\vec{q}\times \vec{p}_{\mu}|}\,,
 \label{eqn:pt}
\ee 
respectively, and 
\be
\rho_0(x,y)&=&\frac{1}{2}e^2 G^2_F sin^2\theta_c 
(1-\lambda)\left\{ {4m^2_{\mu}|f_K|^2
\over \lambda x^2}\left[x^2+2(1-r_{\mu})\left(1-x-{r_{\mu}\over 
\lambda}\right)\right]\right. 
\nonumber\\
&&+ m^4_Kx^2\left[|F_V+F_A|^2{\lambda^2 \over 1-\lambda}
\left(1-x-{r_{\mu}\over \lambda}\right)+
|F_V-F_A|^2(y-\lambda)\right]
\nonumber \\
&& 
-4m_K m^2_{\mu}\left[Re[f_K(F_V+F_A)^*]\left(1-x-{r_{\mu}\over 
\lambda}\right) \right. 
\nonumber \\
&&- \left. \left. Re[f_K(F_V-F_A)^*]{1-y+\lambda \over 
\lambda}\right]\right\} 
\ee
with
$\lambda=(x+y-1-r_{\mu})/x$, $r_{\mu}=m^2_{\mu}/M^2_K$
and $x=2p\cdot q/p^2=2E_{\gamma}/m_K$ and $y=2p\cdot 
p_{\mu}/p^2=2E_{\mu}/m_K$ 
being normalized energies of the photon and muon, respectively.
If we define the longitudinal, normal and transverse, 
muon polarization asymmetries by
\be
P_i(x,y)={d\Gamma(\vec{e}_i)-d\Gamma(-\vec{e}_i) \over 
d\Gamma(\vec{e}_i)+d\Gamma(-\vec{e}_i)}\,, \ (i=L,N,T)\,,
\label{eqn:Pi}
\ee
we find that
\be
P_i(x,y)={\rho_i(x,y)\over \rho_0(x,y)}\,, \ (i=L,N,T)\,,
\label{eqn:pt0}
\ee
with
\be
\rho_L(x,y) &=&
-e^{2}G_{F}^{2}\sin ^{2}\theta _{c}{(1-\lambda)\over 2\lambda 
\sqrt{y^{2}-4r_{\mu}}} 
\left\{ 
{4m_{\mu }^{2}|f_{K}|\over \lambda x^{2}}
\left[ 
x(\lambda y-2r_{\mu})(x+y-2\lambda) 
\right.\right.
\nn\\
&&
\left.
-(y^{2}-4r_{\mu})(\lambda x+2r_{\mu}-2\lambda)
\right]
-M_{K}^{4}\lambda x^{2}
\left[ 
|V+A|^{2}{\lambda \over 1-\lambda}
(\lambda y-2r_{\mu })  \right.
\nn\\
&&
\left.
\left( 
1-x-{ r_{\mu}\over \lambda }
\right)
+| V-A| ^{2}
\left(
y^{2}-\lambda y-2r_{\mu}
\right) \right]
\nn\\
&&-4M_{K}m_{\mu }^{2}\left[Re\{f_{K}( V+A)^{*}\}\lambda \left(
1-x-\frac{r_{\mu }}{\lambda }\right)(2-2x-y) \right.
\nn
\\
&&\left.\left.
+Re\{f_{K}(V-A)^{*}\}\left(( 1-y) 
(y-\lambda) +2r_{\mu }-\lambda \right) \right]\right\}\,,
\nn\\
\rho_N(x,y) &=&
e^{2}G_{F}^{2}\sin ^{2}\theta _{c}{(1-\lambda) 
\sqrt{\lambda y-\lambda ^{2}-r_{\mu }}
\over M_{K}\lambda \sqrt{y^{2}-4r_{\mu }}}
\left\{{4m_{\mu }^{3}|f_{K}| ^{2} \over \lambda x}(x+y-2\lambda)  
\right.
\nn
\\
&&-M_{K}^{4}m_{\mu }\lambda x^{2}\left[ |V+A| ^{2}{\lambda\over 
1-\lambda }\left( 1-x-\frac{r_{\mu }}{\lambda }\right) +|V-A|^{2}
\right]  
\nn
\\
&&-2M_{K}^{3}m_{\mu }\left[Re\{f_{K}(V+A)^{*}\}\left( 
{(r_{\mu }-\lambda )(1-x-r_{\mu })\over 1-\lambda }+\lambda x(1-x)\right)  
\right.
\nn\\
&&\left.\left.
-Re\{f_{K}( V-A)^{*}\}(y-2r_{\mu })\right]\right\}\,,
\nn\\
\rho_T(x,y)&=&-2e^2G^2_F sin^2\theta_c m^2_K m_{\mu}{1-\lambda \over \lambda}
\sqrt{\lambda y-\lambda^2-r_{\mu}}\left\{ Im[f_K(F_V+F_A)^*]{\lambda \over 
1-\lambda}\right.
\nonumber \\
&& \left. \times\left(1-x-{r_{\mu}\over \lambda}\right)+
Im[f_K(F_V-F_A)^*]\right\}\,.
\ee
 From Eq. (\ref{eqn:Pi}), it is easily seen that
the asymmetries of $P_L$ and $P_N$ are even quantities under time-reversal 
transformation, while $P_T$ is an odd one.
Since we are interested in CP violation, we will concentrated on the 
transverse part of the polarization shown in Eq. (\ref{eqn:Pi}).
We rewrite $P_L(x,y)$ as
\be
P_T(x,y) &=& P^V_T(x,y)+P^A_T(x,y)
\label{eqn:pt1}
\ee
with
\be
P^V_T(x,y)&=&\sigma_V(x,y)[Im(\Delta_A+\Delta_V)]\,,
\nn\\
P^A_T(x,y)&=&[\sigma_V(x,y)-\sigma_A(x,y)]Im(\Delta_P)\,,
\label{eqn:pva}
\ee
where
\be
\sigma_V(x,y)=2e^2G^2_F sin^2\theta_c m^2_K m_{\mu}f^0_KF^0_V
{\sqrt{\lambda y-\lambda^2-r_{\mu}}\over \rho_0(x,y)}
\left({-1+\lambda \over \lambda}-
\left(1-x-{r_{\mu}\over \lambda}\right)\right),
\nn\\
\sigma_A(x,y)=2e^2G^2_F sin^2\theta_c m^2_K m_{\mu}f^0_K F^0_A
{\sqrt{\lambda y-\lambda^2-r_{\mu}}\over \rho_0(x,y)}
\left({-1+\lambda\over \lambda}+
\left(1-x-{r_{\mu}\over \lambda}\right)\right).
\ee
Clearly, to have CP violating transverse muon polarization of $P_T$ in
Eq. (\ref{eqn:pva}), at least one of the couplings $G_i\ (i=P,A,V)$ in
Eq. (2) has to exist and be complex.

In the standard model, from Eqs. (\ref{eqn:pt1}) and (\ref{eqn:pva}),
we see that the transverse muon 
polarization is 
zero since $f_K=f_K^0$ and $F_{V(A)}=F_{V(A)}^0$ which are both real 
due to $G_i=0$ and $\Delta_i=0\ (i=P,V,A)$. 
However, the longitudinal and normal muon polarizations are 
non-vanishing.
In Figs. 1-3, we show the Dalitz plots for $\rho_0$, $P_L$ and $P_N$, 
respectively.
In the non-standard models, $P_T$ could be non-zero if the new physics 
couplings $G_i\ (i=P,A,V)$ have some phases.
In Figs. 4 and 5, we display
the Dalitz plots of $\sigma_V(x,y)$ and $\sigma_V(x,y)-\sigma_A(x,y)$, 
respectively. From the figures, we see that they are all in the
order of $10^{-1}$ in most of the allowed parameter space. We shall use
$\sigma_V$ and $\sigma_V-\sigma_A$ as $0.1$ in our numerical estimations of 
the next section. However,
it is interesting to note that 
there is
no contribution to the transverse muon polarization 
if the interaction beyond the standard model contains
only left-handed vector current, $i.e.$, $G_V=-G_A$, 
because of the zero
relative phase between the amplitudes of $M_{IB}$ and $M_{SD}$. 
  
\se{Transverse Muon Polarization in Various Models} 

\ \ \ 
Since $P_T(K^+_{\mu 2\gamma})=0$ at tree level in the standard model, a
nonzero value of the transverse muon polarization provides an evidence for
new CP violating source outside the CKM mechanism after taking care of the
theoretical background. In such case, $P_T$ may arise from the
interference between the tree level amplitude in the standard model and
the new CP violating amplitude.  In the following we will study various CP
violation models such as the left-right symmetric, two-Higgs-doublets,
supersymmetry (SUSY), and leptoquark models
and discuss the possibilities of having large $P_T$ in these theories. 

\sse{Left-Right symmetric models}
\ \ \ 
In this subsection we study the prediction of T-violating muon polarization
for $K^+_{\mu 2\gamma}$ decay in
models with left-right symmetric gauge symmetries $SU(2)_L \times
SU(2)_R \times U(1)_{B-L}$ \cite{Pati}. 
In these models, 
the minimal set of Higgs multiplets to break the gauge symmetry down to 
$U(1)_{em}$ is: one doublet $\phi$ and two triplets $\Delta_{L,R}$.
The transformations of the Higgs bosons under $SU(2)_L \times SU(2)_R $ 
are given by
\be
\phi&=&\left( \begin{array}{c c}
\phi^{0^*}_1 & \phi^+_2 \\
-\phi^-_1 & \phi^+_2  \end{array}\right) \longrightarrow
U_L \, \phi \, U^{\dagger}_R,  \nonumber \\
\Delta_{L,R}&=& \left( \begin{array}{c c}
\delta^+/\sqrt{2} & \delta^{++} \\
\delta^0 & -\delta^+/\sqrt{2} \end{array} \right)_{L,R} \longrightarrow
U_{L,R}\, \Delta_{L,R}\, U^{\dagger}_{L,R}\,,
\ee
with the vacuum expectation values (VEVs) being
\be
<\phi>&=&{e^{i\alpha} \over \sqrt{2}}\left( \begin{array}{c c}
v_1 & 0 \\
0 & v_2 \end{array} \right) \nonumber \\
<\Delta_{L,R}>&=&{e^{i\theta_{L,R}} \over \sqrt{2}} \left( \begin{array}{c c}
0 & 0 \\
v_{L,R} &0 \end{array} \right),
\ee
where $\alpha$ and $\theta_{L,R}$ are the CP violating phases. 
After the spontaneous symmetry breaking (SSB),
the masses of fermions can be generated by the Yukawa coupling terms
\be
{\cal L}_Y= \bar{Q}_L F \phi Q_R + \bar{Q}_L G \tilde{\phi} Q_R +h.c.,
\ee
where the generation indices have been suppressed, 
$Q_{L,R}$ stand for the left- and right-handed fermions doublets,
F and G correspond to the $N\times N$ mass matrices for N generations,
and $\tilde{\phi}\equiv \tau_2 \phi^* \tau_2$. In addition, the eigenstates
of both $W_{L,R}$ bosons related to weak eigenstates can be written as 
\be
W_1=\cos\xi \,W_L + \sin\xi \,W_R, \nonumber \\
W_2=-\sin\xi \,W_L + \cos\xi \,W_R,
\ee
where $\xi$ is the left-right mixing angle.
In any event, due to the $W_R$ gauge boson, we have new mixing matrix,
called right-handed CKM (RCKM) 
matrix, coming from the diagonalization of right-handed quarks.
The charged current interactions
are given by
\be
{\cal L}_{CC}={g_L \over \sqrt{2}}W^{\mu}_L \bar{U}\gamma_{\mu} K^L P_L D+
{g_R \over \sqrt{2}}W^{\mu}_R \bar{U}\gamma_{\mu} K^R P_R D+h.c.,
\ee
where $g_L$ and $g_R$ are coupling constants for $SU(2)_L$ 
and $SU(2)_R$, $U^{T}=(u,c,t)$ and $D^T=(d,s,b)$ are 
the physical states of up-type quarks and down-type quarks, $K^L$ and $K^R$
are the CKM and RCKM matrices, and $P_{L,R}$=$(1\mp \gamma_5)/2$, 
respectively. 

The number of physical CP violating 
phases in CKM and RCKM matrices with N generations
are $N_L=(N-1)(N-2)/2$ and $N_R=N(N+1)/2$, respectively.
For example, we have three CP violating phases from $K^R$
for the case of two generations.
Here, we do not assume parity invariant 
as well as any special relation in the matrices so that, in general, the 
gauge coupling constant $g_L$ is not equal to $g_R$ and the matrix 
elements of RCKM are free parameters. 
Including the left-right mixing parameter $\xi$
we can generate $P_T(K^+_{\mu 2\gamma})$ 
from the tree diagram as shown in Fig. 6. 
The quark level four-Fermion interaction which contributes to 
$K^+_{\mu 2\gamma}$ decay is given by
\be
{\cal L}_{RL}=-2\sqrt{2}G_F\left(\frac{g_R}{g_L}\right)K^{R^*}_{us}\,\xi
\, \bar{s} \gamma_{\mu}P_R\,u\; \bar{\nu}\, \gamma^{\mu}P_L\, \mu.
\label{eqn:25}
\ee
Compare Eq. (\ref{eqn:25}) with Eq. (\ref{eqn:fermi}), we get 
\be
G_V=G_A=-\frac{G_F}{\sqrt{2}} {g_R \over g_L} K^{R^*}_{us} \xi\,,
\ee
which leads to
\be
\Delta_A=\Delta_V=-{ \xi\, K^{R^*}_{us}\over sin\theta_c}
{g_R \over g_L}\,,
\label{eqn:lrav}
\ee
in terms of Eq. (\ref{eqn:fk}).
 From the expression of $\delta_{A,V}$ in Eq. (\ref{eqn:lrav}) and the 
definition of the muon transverse polarization in Eq. ({\ref{eqn:pt1}),
we obtain
\be
P_T&=& 2\sigma_V { \xi g_R \over g_L}Im(K^{R^*}_{us})\,.
\label{eqn:lrp}
\ee
To illustrate the prediction of the transverse polarization,
we consider 
two-generation case.
The RCKM matrix which contains three physical phases can be parametrized as
\be
U_R=e^{i\gamma} \left [\begin{array}{cc}
e^{-i\delta_2} cos\theta_c & e^{-i\delta_1} sin\theta_c \\
-e^{i\delta_1} sin\theta_c & e^{i\delta_2} cos\theta_c \end{array} \right ],
\label{eqn:29}
\ee 
$\delta_{1,2}$ and $\gamma$ are the three physical CP violating 
phases, and $\theta_c$ is the Cabibbo angle. 
From the mixing matrix in Eq. (\ref{eqn:29}), we find that
\be
Im(K^R_{us})=sin\theta_c\, 
sin(\gamma-\delta_1)\leq sin\theta_c\,.
\ee

In the typical left-right symmetric models as shown in Ref. \cite{Wolf}, one 
generally has 
\be
\xi {g_R\over g_L} &<&  4.0 \times 10^{-3}\,,
\ee
and thus one gets
\be
 P_T&<& 8.0\times 10^{-4}\,
\label{eqn:ptlr1}
\ee
with choosing that $\sigma_V\sim 0.1$ and $Im(K^{R^*}_{us})\sim \sin\theta_c$.
However, in a class of the specific 
models studied in Ref. \cite{BBPR}, it is found 
that 
\be
\xi {g_R\over g_L} < 3.3 \times 10^{-2} 
\ee
for $M_R>549$ GeV.
In such models, with the same set of the parameters $\sigma_V$ and
$Im(K^{R^*}_{us})$, one gets
\be
 P_T&<& 6.6\times 10^{-3}\,,
\label{eqn:ptlr2}
\ee
which is within the experimental detecting range.
The bounds in Eqs. (\ref{eqn:ptlr1}) and (\ref{eqn:ptlr2}) can be 
even larger if one uses a larger value of $\sigma_V$.
We remark that
the muon transverse polarization for the decay of $K^+\to \pi^0 \mu^+ \nu$,
$i.e$., $P_{T}(K^+_{\mu 3})$, vanishes in the left-right symmetric models
as shown in Ref. \cite{BG}. This is because the photon is a vector 
particle while the pion is a pseudo-scalar one.

\subsection {Two-Higgs-Doublet Models with FCNC}
\ \ \  
In a two-Higgs doublet model (THDM) without introducing 
global symmetry \cite{Lee}, the up and down-type quarks will 
couple to the both two Higgs doublets. However, 
the up and down-type quarks mass matrices cannot be
simultaneously diagonalized.
Therefore, flavor changing neutral current (FCNC)
is induced at tree level. 
To suppress the effects of the FCNC, one can impose a discrete symmetry.
However, the discrete symmetry normally also constrains the scalar potential 
such that spontaneous CP violation (SCPV) may not occur. 
In this section, instead of having a discrete symmetry, we 
assume that the couplings related to the FCNC are small.
The various possible theories with naturally small couplings have been 
explored in Refs. \cite{Wu,Hall}.

The Yukawa coupling terms in the weak eigenstate can be written by
\be
{\cal L}_Y&=&\eta^D_{ij}\bar{Q}_{Li} \phi_1 D_{Rj} + \eta^U_{ij}\bar{Q}_{Li}
\tilde{\phi}_1 U_{Rj} + \xi^D_{ij}\bar{Q}_{Li} \phi_2 D_{Rj} + 
\xi^U_{ij}\bar{Q}_{Lia} \tilde{\phi}_2 U_{Rj}+\nonumber \\
&&\eta^E_{ij}\bar{L}_i \phi_1 E_{Rj}+\xi^E_{ij}\bar{L}_{i} \phi_2 E_{Rj}+h.c. 
,
\label{eqn:ly}
\ee
where $\eta^{U,D,E}_{ij}$ and $\xi^{U,D,E}_{ij}$ are dimensionless and
real parameters, $Q_L$ and L denote the left-handed quark and 
lepton doublets,
$D_R$, $U_R$, and $E_R$ are the right-handed 
down-type quarks, up-type quarks, and leptons, 
$\phi_1$ and $\phi_2$ are 
the two Higgs doublets with $\tilde{\phi}\equiv i\sigma_2\phi^*$
respectively. 
The VEVs of the Higgs doublets are given by $<\phi_1>=v_1$ and 
$<\phi_2>=exp(i\theta)v_2$, where $\theta$ is the CP violating phase. 
 From Eq. (\ref{eqn:ly}) the Yukawa 
interactions of quarks and leptons with neutral and charged Higgs bosons 
can be expressed by
\be
\nn
{\cal L}_{NH}&=&\frac{1}{\sqrt{2}}\bar{U}_L \tilde{\xi}^U U_R (h^0-iA^0)+
\frac{1}{\sqrt{2}}\bar{D}_L \tilde{\xi}^D D_R (h^0+iA^0)+ h.c.,
\label{eqn:nh} \\
{\cal L}_{CH}&=&-\bar{U}_R \tilde{\xi}^{U^{\dagger}} K^L D_L H^{\dagger}+
\bar{U}_L K^L \tilde{\xi}^D D_R H^{\dagger} +
\bar{N}_L K^L \tilde{\xi}^E E_R H^{\dagger} + h.c.\,, 
\label{eqn:ch}               
\ee 
respectively, with 
\be
\nn
\tilde{\xi}^U&\equiv& V^U_L(-\eta^U \sin\beta\, + e^{-i\theta} \xi^U 
\cos\beta\,) {V^U_R}^{\dagger},\\ 
\nn
\tilde{\xi}^D&\equiv& V^D_L(-\eta^D \sin\beta\, + e^{i\theta} \xi^D
\cos\beta\,) {V^D_R}^{\dagger},\\  
\tilde{\xi}^E&\equiv& (-\eta^E \sin\beta\, + e^{i\theta} \xi^E
\cos\beta\,) {V^E_R}^{\dagger},    
\ee
where flavor indices are suppressed,
$V^{U,D,E}_{L,R}$ are the unitary matrices which transform the fermionic
weak-eigenstates to the mass-eigenstates, and $\tan\beta=v_2/v_1$.
However, for simplicity we can reparameterize the diagonal parts of matrix
$\tilde{\xi}^E$ to be $(\tilde{\xi}^E)_{ii}\equiv z m_{l_{i}}$,
where $z$ is unknown parameter and $m_l$ is the lepton mass. The 
parameter $z$ can be bounded by the $\mu$-e universality in tau decay.

 From Eq. (\ref{eqn:ch}), 
we find the following four-Fermion interaction
\be
{\cal L}_{K^+_{\mu 2}}={z m_{\mu} \over M^2_H} \bar{s}\left (-\sum_j 
K^{L^*}_{js} \tilde{\xi}^U_{j1} P_R + \sum_{i} \tilde{\xi}^{D^*}_{i2} 
K^{L^*}_{ui} P_L \right )u \bar{\nu} P_R \mu.
\label{eqn:k2}
\ee
where $M_H$ is the mass of charged Higgs particle.
 From Eq. (\ref{eqn:nh}), we clearly see that the off-diagonal elements 
$\tilde{\xi}^{U,D}_{ij}\ (i\neq j)$ 
are related to the FCNC at tree level.
To illustrate the polarization effect, we use 
the ansatz in Ref. \cite{Sher}
by Cheng and Sher,
 in which the couplings
$\tilde{\xi}^{U,D}$ in Eq. (\ref{eqn:k2}) are taken to be
\be
\tilde{\xi}^{U,D}_{ij}=\lambda_{ij}{\sqrt{m^{U,D}_i m^{U,D}_j} \over v}
\hspace{2cm} {\rm for} \ \ i\neq j,
\label{eqn:ansa}
\ee
where $\lambda_{ij}$ are undetermined parameters, $v=\sqrt{v^2_1+v^2_2}=
(2\sqrt{2} G_F)^{-1/2}$, and $m^{U(D)}$ the masses of
up (down)-type quarks. 
We now simplify Eq. (\ref{eqn:k2}) to
\be
{\cal L}_{K^+_{\mu 2}}=
{z m_{\mu} \over M^2_H}\bar{s}K^{L^*}_{us}(-\tilde{\xi}^U_{11}P_R+
\tilde{\xi}^{D^*}_{22} P_L)u\; \bar{\nu} P_R \mu\,,
\label{eqn:k2-1}
\ee
by using Eq. (\ref{eqn:ansa}).
We consider the constraints on $\tilde{\xi}^U_{11}$ and 
$\tilde{\xi}^D_{22}$ from the $K^0-\bar{K}^0$ mixing, arising from the 
box diagrams with the $u$-quark as the internal fermion as shown in Fig. 7.
 From the figures, we find that
\be
\nn
<K^0|M^{box}_{HH}|\bar{K}^0>&=&-{f^2_K m_K \over 6 (4\pi)^2 
M^2_H}(K^L_{ud}
K^L_{us})^2(\tilde{\xi}^D_{11}\tilde{\xi}^{D^*}_{22}-\tilde{\xi}^U_{11}
\tilde{\xi}^{U^*}_{11})^2, \label{eqn:mhh}\\
<K^0|M^{box}_{HW}|\bar{K}^0>&=&{4\sqrt{2}G_F \over (4\pi)^2}
{f^2_k m^3_K \over 3 m^2_s}
(K^L_{ud}K^L_{us})^2 \tilde{\xi}^{D}_{11} \tilde{\xi}^{D^*}_{22}
D_{HW}(\frac{M^2_H}{M^2_W},\frac{m^2_u}{M^2_W})
\label{eqn:mhw}
\ee 
with 
\be
D_{HW}(x,y)={\ln x \over 2(x-1)x^2}+{\ln x \over 2x}+{\ln(xy) \over 2x^2}.
\ee
 Using the experimental value on the 
$K^0-\bar{K}^0$ mixing, we obtain
\be
\nn
|\tilde{\xi}^D_{11}\tilde{\xi}^{D^*}_{22}-\tilde{\xi}^U_{11}
\tilde{\xi}^{U^*}_{11}|&<& 7.3 \times 10^{-5} M_H \, {\rm GeV^{-1}},
\\
|\tilde{\xi}^D_{11} \tilde{\xi}^{D^*}_{22}|
&<&1.3\times 10^{-2} m_s{{\rm GeV^{-1}}\over \sqrt{D_{HW}
(\frac{M^2_H}{M^2_W},\frac{m^2_u}{M^2_W})}}.
\ee 
For $M_H\sim 200\ GeV$, we find that
$|\tilde{\xi}^D_{11}\tilde{\xi}^{D^*}_{22}|<5.6\times 10^{-3}$ and thus,
\be
|\tilde{\xi}^U_{11}|<8.6\times 10^{-3}\sqrt{M_H}.
\label{eqn:c1}
\ee
The strict constraint on $z$, as pointed out by Grossman \cite{Grossman},
is from the $\mu$-e universality in $\tau$ decay as well as the 
perturbativity bound, and is given by
\be
|z|<min(1.1\times 10^{-2} M_H\, GeV^{-2}, 2.0 GeV^{-1})\,.
\label{eqn:39}
\ee 
One notes that 
if $M_H>175\ GeV$,
the bound on $z$ mainly comes from 
the perturbativity as shown in Eq. (\ref{eqn:39}).
 From Eqs. (\ref{eqn:fermi}), (6) and 
(\ref{eqn:k2-1}), we get 
\be
G_P=-{z m_{\mu} \over 4 M^2_{H}} K^{L^*}_{us}
(\tilde{\xi}^U_{11}+\tilde{\xi}^{D^*}_{22})\,, 
\ee
and
\be
\Delta_P=-{m^2_K \over m_s+m_u}
{\sqrt{2}\, z \over 4G_F  M^2_H}
 ( \tilde{\xi}^U_{11}+
 \tilde{\xi}^{D^*}_{22})\,,
\ee
which leads to the muon transverse 
polarization as
\be
P_T=(\sigma_V-\sigma_A){m^2_K\over m_s+m_u}{\sqrt{2} z\over 4G_F M^2_H}
Im(\tilde{\xi}^U_{11}+
 \tilde{\xi}^{D^*}_{22}).
\ee
Using the constraints in Eqs. (\ref{eqn:c1}) and (\ref{eqn:39}), we obtain
  \be
   P_T\le 0.048
   \ee
for $|\sigma_V-\sigma_A|\sim 0.1$ and $M_H\sim 100$ GeV.

Using the interaction in Eq. (\ref{eqn:k2-1}) by neglecting the contribution
of $\tilde{\xi}^{D^*}_{22}$, the similar estimation for the transverse 
muon polarization in $K^+\to \pi^0 \mu^+ \nu$ can be done.
By taking the same values of parameters as in the decay of $K^+_{\mu  
2\gamma}$, we find that $P_T(K^+_{\mu 3})$ 
could be as large as the current experimental limit, $i.e.$, $1\times 
10^{-2}$. Therefore, $P_T$ in both decays of $K^+_{\mu 2\gamma}$
and $K^+_{\mu 3}$ can be very large in the Higgs models with FCNC.
We note that the muon polarization effects of the two modes in the three 
Higgs-doublet models with NFC could be also large as studied in Refs. 
\cite{Kobayashi,BG,Garisto}. 

\sse {Supersymmetric Models}
\ \ \ 
In this subsection we consider the effects on $P_T$ in theories with
SUSY. It is known that, in general, SUSY theories
would contain couplings with the violation of baryon or/and lepton 
numbers, that could induce the rapid proton decay.
 To avoid such couplings, one usually 
assigns R-parity, defined by $R\equiv (-1)^{3B+L+2S}$ 
to each field \cite{Fayet},
where $B(L)$ and $S$ denote the baryon (lepton) number and 
the spin, respectively.
Thus, the R-parity can be used to distinguish the particle (R=+1) from its 
superpartner (R=$-1$).
 Recently, Ng and Wu \cite{Ng} have investigated the $T$ violating $P_T$ 
for $K^+_{\mu 2\gamma}$ in SUSY models with R-parity and they find that when 
the squark family mixings are taken into account, large enhancement 
effects would appear due to the heavy quark masses and large 
tan$\beta$=$v_2/v_1$. In this paper, we will concentrate on the SUSY models 
without R-parity. 
To evade the stringent constraint from proton decay,
we can simply require that 
B violating couplings do not coexist
with the L violating ones. 
In the following we consider the theories with the violation of 
the R-parity and the lepton number.
In such cases, the superpotential is given by 
\be
W_{\not\!L}=\frac{1}{2}\lambda_{ijk}L_iL_jE^c_k+\lambda'_{ijk}L_iQ_jD^c_k,
\ee
where the subscript $ijk$ are the generation indices, $L$ and $E^c$ 
denote the chiral superfields of lepton doublets and 
singlets, and Q and $D^c$ are the chiral superfields of quark doublets and 
down-type quark singlets, respectively. 
We note that the first two generation indices of
$\lambda_{ijk}$ are antisymmetry, $i.e.$,
$\lambda_{ijk}=-\lambda_{jik}$.   
The corresponding Lagrangian is 
\be
{\cal L}_{\not\! L}&=&\frac{1}{2}\lambda_{ijk}[\bar{\nu}^c_{Li} e_{Lj}
\tilde{e}^*_{Rk}+\bar{e}_{Rk} \nu_{Li} \tilde{e}_{Lj}+
\bar{e}_{Rk}e_{Lj}\tilde{\nu}_{Li}-(i\leftrightarrow j)]
+ \lambda'_{ijk}[\bar{\nu}^c_{Li} d_{Lj}\tilde{d}^*_{Rk} \nonumber \\
&+&\bar{d}_{Rk}\nu_{Li} \tilde{d}_{Rk} 
+\bar{d}_{Rk}d_{Lj}\tilde{\nu}_{Li}- \bar{e}^c_{Ri}u_{Lj}
\tilde{d}^*_{Rk}-\bar{d}_{Rk}e_{Li}\tilde{u}_{Lj}-\bar{d}_{Rk}u_{Lj}
\tilde{e}_{Li}]+h.c..
\label{eqn:lv} 
\ee 
 From Eq. ({\ref{eqn:lv}), we find that the four-Fermion interaction 
of $\bar{s}u\to\mu\bar{\nu}$ with the slepton as the intermediate state 
shown in Fig. 8 can be written as
\be
{\cal L}_{RV}=-{\lambda^*_{2i2}\lambda'_{i12}\over M^2_{\tilde{e}_{Li}}}
\bar{s}\,P_L\, u \bar{\nu}\, P_R \, \mu,
\label{eqn:rv}
\ee
where $M^2_{\tilde{e}_{Li}}$ is the slepton mass.

 From the interaction in Eq. (\ref{eqn:rv}), we get
\be
G_P={\lambda^*_{2i2} \lambda'_{i12} \over 4 M^2_{\tilde{e}_{Li}}}\,,
\ee
which leads to
\be
\Delta_P={\sqrt{2}\over 4G_F sin\theta_c}{m^2_K
\over (m_s+m_u)m_{\mu}}
{\lambda^*_{2i2}\lambda'_{i12}\over M^2_{\tilde{e}_{Li}}}
\ee
for $i=1,3$.
We therefore obtain the transverse muon polarization of 
$K^+_{\mu 2\gamma}$ as
\be
 P_T&=&-(\sigma_V-\sigma_A){\sqrt{2}\over 4G_F\sin\theta_C}{m_K^2\over 
(m_s+m_u)m_{\mu}}{ Im(\lambda^*_{2i2} \lambda'_{i12})\over 
M^2_{\tilde{e}_{Li}}}\,.
\label{eqn:susypt}
 \ee
In order to give the bounds on the R-parity violating couplings, we 
need to examine the various processes induced by the FCNC. 
We first study the decay of $\mu \to e \gamma$. 
Based on the analysis in Ref. \cite{Carlos},
the total branching ratio for $\mu \to e \gamma$ decay can be expressed by
\be
Br(\mu \to e \gamma)={12\pi^2\over G_F} (|A_{LR}|^2+|A_{RL}|^2),
\ee  
where
\be
A_i=A^{\lambda}_i+A^{\lambda'}_i+A^{\Delta m}_i
\label{eqn:50}
\ee
with $i=LR$ and $RL$. In Eq. (\ref{eqn:50}), the amplitudes $A^{\Delta 
m}_i$ stand for the 
neutralino- and slepton-mediated contributions and $\Delta m$ is the soft
breaking mass. For convenience, we only consider 
$A^{\lambda}_i$ contributions. Using the results in
Ref. \cite{Carlos}, we get 
\be
A^{\lambda}_{LR}&=&{e\over 96\pi^2}\sum^3_{i,k=1}\lambda_{i1k}\lambda_{i2k}
\left( {1\over M^2_{\tilde{\nu}_i}}-\frac{1}{2}{1\over M^2_{\tilde{e}_{Rk}}}
\right),
\nonumber \\
A^{\lambda}_{RL}&=&{e\over 96\pi^2}\sum^3_{i,j=1}\lambda_{ij1}\lambda_{ij2}
\left( {1\over M^2_{\tilde{\nu}_i}}-\frac{1}{2}{1\over M^2_{\tilde{e}_{Lj}}}
\right).
\ee
Therefore, the bounds on R-parity violating couplings $\lambda$ 
are given by
\be
{|\lambda_{31k}\lambda_{32k}|\over M^2}&<&4.6\times 10^{-8}{1\over 
GeV^2}, \hspace{2cm} {\rm for} \ k=1,2;\\
\nn
{|\lambda_{ij1}\lambda_{ij2}|\over M^2}&<&2.3\times 10^{-8}{1\over 
GeV^2}, \hspace{2cm} {\rm for} \ i,j=1,2,
\ee
where we have assumed that $M_{\tilde{\nu}_{\tau}}\simeq M_{\tilde{e}_R}
\simeq M_{\tilde{e}_L}\simeq M$.
 For simplicity, we take $|\lambda_{31k}|\sim |\lambda_{32k}|$ and 
$|\lambda_{211}|\sim |\lambda_{212}|$, and we have
\be
{|\lambda_{322}|\over M}&<&2.1\times \times 10^{-4}{1\over 
GeV},\nonumber \\
{|\lambda_{212}|\over M}&<&1.5\times 10^{-4}{1\over GeV}\,.
\label{eqn:b1}
\ee
The bounds on $\lambda'$ can be extracted from the experimental limit on
$K^+\to \pi \nu \bar{\nu}$ as shown in Ref. \cite{Agashe}. One finds that
\be
{|\lambda'_{ijk}|\over M_{\tilde{d}_{Rk}}}
<1.2\times 10^{-4}{1\over GeV}\,,
\hspace{2cm} {\rm for} \ j=1,2,
\label{eqn:b2}
\ee
where $M_{\tilde{d}_{Rk}}\sim M$ is the sdown-quark mass. 

 From Eq. (\ref{eqn:susypt}) and the bounds in Eqs. (\ref{eqn:b1}) and
(\ref{eqn:b2}), we find
\be
 P_T(K^+_{\mu 2\gamma})\le 0.01
\ee
for $|\sigma_V-\sigma_A|\sim 0.1$.

The interaction in Eq. (\ref{eqn:rv}) could also yield transverse muon 
polarization in $K^+\to\pi^0\mu^+\nu$. We find \cite{BG} that
$P_T(K^+_{\mu 3})< 10^{-2}$
by using the same parameter values as in the case 
of $K^+_{\mu 2\gamma}$.
Thus, in R-parity violation SUSY models with R-parity violation, one can 
also get a large prediction of $P_T(K^+_{\mu 3})$.

\sse{Leptoquark Model}
\ \ \
There exist three scalar leptoquark models contributing to the decay
$K^+_{\mu 2\gamma}$ through the tree diagrams which is similar to the 
cases for $K^+_{\mu 3}$ shown in Ref. \cite{BG}.
 The quantum numbers of the
leptoquarks under the standard group
$SU(3)_C\times SU(2)_L\times U(1)_Y$ are \cite{BG,Nieves,DH} 
\be
\nn
\phi_1&=&(3,2,{7\over 3})\,,\ (Model\ I),\\
\nn
\phi_2&=&(3,1,-{2\over 3})\,,\ (Model\ II),\\
\phi_3&=&(3,3,-{2\over 3})\,,\ (Model\ III),
\label{eqn:lq}
\ee
respectively. The general
couplings involving these leptoquarks are given by \cite{DH}
\be
\nn
{\cal L}^I & = & (\lambda_1\bar{ Q}_Le_R+\lambda_1'\bar{u}_RL_L)\phi_1
+h.c., \label{eqn:l1}\\
\nn
{\cal L}^{II} & = & 
(\lambda_2\bar{Q}_LL_L^c+
\lambda_2'\bar{ u}_Re_R^c)\phi_2+H.C.,\label{eqn:l2}\\
{\cal L}^{III} & = & \lambda_3\bar{Q}_LL_L^c\phi_3+H.C.,
\label{eqn:lqint}
\ee
where $Q=\left(\begin{array}{c} u\\d \end{array}\right)$ and
$L=\left(\begin{array}{c} \nu\\e \end{array}\right)$.
Here the coupling constants $\lambda_k\ (k=1,\ldots ,3)$ are complex 
and thus CP violation could arise from the Yukawa interactions in 
Eq. (\ref{eqn:lqint}).
We assume that CP violation in $K\to \pi\pi$ decays
can be accounted for by the non-vanishing KM phase, and
investigate the effect on the muon polarization
of adding another CP violation mechanism in 
Eq. (\ref{eqn:lqint}).
 
In terms of each charge components of the leptoquarks we rewrite 
Eq. (\ref{eqn:lqint}) as
\be
{\cal L}^{I} & = &\sum_{i,j}\{[\lambda_1^{ij}\bar{u}_i{1\over 2}(1+\gamma_5)e_j+
{\lambda'}_1^{ij}\bar{u}_i{1\over 2}(1-\gamma_5)e_j]\phi_1^{({5\over 3})}
    +\nonumber\\
\nn
 &  &[\lambda_1^{ij}\bar{d}_i{1\over 2}(1+\gamma_5)e_j+
{\lambda'}_1^{ij}\bar{u}_i{1\over 2}(1-\gamma_5)\nu_j]\phi_1^{({2\over 3})}\}+
H.C.,
\\
{\cal L}^{II} & = &\sum_{i,j}\{
{\lambda}_2^{ij}[-\bar{u}_i{1\over 2}(1+\gamma_5)e_j^c
    +\nonumber\\
\nn
 &  &\bar{d}_i{1\over 2}(1+\gamma_5)\nu_j^c]+
{\lambda'}_2^{ij}\bar{u}_i{1\over 2}(1-\gamma_5)e_j^c\}\phi_2^{(-{1\over 3})}+
H.C.,\\
{\cal L}^{III} & = &\sum_{i,j}
\lambda_3^{ij}\{\bar{u}_i{1\over 2}(1+\gamma_5)\nu_ j^c
\phi_3^{({2\over 3})}+
[\bar{u}_i{1\over 2}(1+\gamma_5)e_j^c    +\nonumber\\
 &  &\bar{d}_i{1\over 2}(1+\gamma_5)\nu_j^c]\phi_3^{(-{1\over 3})}
+
\bar{d}_i{1\over 2}(1+\gamma_5)e_j^c\phi_3^{(-{4\over 3})}\}+H.C.,
\label{eqn:LIII}
\ee
where $i,j$ are family indices and $Q_e$ in $\phi_k^{(Q_e)}$ are the
electric charges. From 
Eq. (\ref{eqn:LIII}),
 we see that the relevant terms for the
process $K^+_{\mu 2\gamma}$ 
are the ones involving $\phi_1^{({2\over 3})}\,,\
\phi_2^{(-{1\over 3})}$ and $\phi_3^{(-{1\over 3})}$ couplings, respectively.
We will concentrate on these terms in our discussions. The effective
interactions from these leptoquark exchanges  are
\be
\nn
{\cal L}^I_{eff} & = & {\lambda_1^{22}({\lambda'}_1^{1i})^{*}\over 4M^2_{\phi_1}
}
\bar{s}(1+\gamma_5)\mu\bar{\nu}_i(1+\gamma_5)u+H.c\,, \\
\nn
{\cal L}^{II}_{eff} & = & {1\over 4M^2_{\phi_2}}
[
-{\lambda}_2^{2i}({\lambda}_2^{12})^{*}
\bar{s}(1+\gamma_5)\nu_i^c\bar{\mu}^c(1-\gamma_5)u+\nonumber\\
&&
{\lambda}_2^{2i}({\lambda'}_2^{12})^{*}
\bar{s}(1+\gamma_5)\nu_i^c\bar{\mu}^c(1+\gamma_5)u]+H.C.\,,\\
{\cal L}^{III}_{eff} & = & 
{\lambda_3^{2i}(\lambda_3^{12})^{*}\over 4M^2_{\phi_3}
}
\bar{s}(1+\gamma_5)\nu_i^c\bar{\mu}^c(1-\gamma_5)u+H.C
\label{eqn:eff}
\ee
where $M_{\phi_k}\ (k=1,\ldots ,3)$ are the masses of
$\phi_1^{({2\over 3})}\,,\
\phi_2^{(-{1\over 3})}$ and $\phi_3^{(-{1\over 3})}$, respectively.
Using the Fierz transformations
and Eq. (\ref{eqn:fkp}), we have
\be
\nn
\Delta^I_P & = &- {\sqrt{2}m^2_K\over G_F sin\theta_c (m_s+m_u)m_{\mu}}
{\lambda^{22}_1(\lambda'^{1i}_1)^*\over 8M^2_{\phi_1}},
\\ 
\Delta^{II}_P & = &- {\sqrt{2}m^2_K\over G_F sin\theta_c (m_s+m_u)m_{\mu}}
{\lambda^{2i}_2(\lambda'^{12}_2)^*\over 8M^2_{\phi_2}},
\label{eqn:delta2}
\ee
which give rise to
\be
[P^I_T,P^{II}_T]=(\sigma_V-\sigma_A){\sqrt{2}\over G_F 
sin\theta_c}{m^2_k\over (m_s+m_u)m_{\mu}}
\left[{\lambda^{22}_1 (\lambda'^{1i}_1)^* \over M^2_{\phi_2}}, 
{\lambda^{2i}_2(\lambda'^{12}_2)^*\over 8M^2_{\phi_2}}\right]\,,
\ee
respectively,
where we have neglected the tensor interactions and the contribution of 
${\cal L}^{III}$ 
because it contains only  left-handed vector current interactions which 
have no relative phases between $M_{IB}$ and $M_{SD}$.  

For Model I, since the leptoquarks model can give a large contribution 
to the moun transverse polarization in $K^+_{\mu 3}$ \cite{BG}, we 
may use the present experimental bound on $P_T(K^+_{\mu 3})$ to 
constrain the CP violating parameters in Eq. (\ref{eqn:eff}).
Explicitly, we find 
\be
{|\lambda^{22}_1 (\lambda'^{1i}_1)^*| \over M^2_{\phi_1}}<4.4\times 10^{-8},
\ee 
and therefore we get
\be
|P^{I}_T(K^+_{\mu 2 \gamma})|<4.8\times 10^{-3}
\ee 
for $|\sigma_V-\sigma_A|\sim 0.1$. 
Similar results can be obtained in Model II.

\se {Conclusions}

\ \ \
We have studied the transverse muon polarization
in the decay of $K^+\to \mu^+\nu\gamma$ in various CP violation theories.
We have explicitly demonstrated that the polarization effect can be large
in models with the left-right symmetry, multi-Higgs bosons, SUSY, and 
leptoquarks, repectively.  These results 
as well as that from other CP violation models are summarized in Table 1. 
The estimations for the transverse muon polarization in $K^+_{\mu 3}$ are 
also presented in Table 1. 
It is interesting to see that
a large $P_T(K^+_{\mu 2\gamma})$ corresponds a large $P_T(K^+_{\mu 3})$ 
in the most CP violation models shown in the table except the
left-right symmetric theories, in which $P_T(K^+_{\mu 3})=0$.
Therefore, the decay of $K^+_{\mu 2\gamma}$ has a comparable or even more 
sensitivity with that of $K^+_{\mu 3}$ to the new CP violation mechanism.

In conclusion, the transverse muon polarization of 
$K^+\to \mu^+\nu\gamma$ could be at the level of $10^{-2}$
in the non-standard CP violation theories,
which may be detectable at the ongoing KEK
expeiment of E246 as well as the proposed BNL experiment.
The measurement of such effect
is a clean signature of CP violation beyond the standard model since that 
from the theoretical background, $i.e.$, FSI, is  $\le 10^{-3}$.

\vspace{1cm}

\noindent
{\bf Acknowledgments}

This work is supported by the National Science Council of the
ROC under contract numbers NSC86-2112-M-007-021 and 
NCHC-86-02-007.

\newpage
{\bf Table 1.}\  Summary of the upper values of
(1) $P_T(K^+_{\mu 3})$ and (2) $P_T(K^+_{\mu 2\gamma})$ for
(a) current
experimental limits, (b) sensitiveties in KEK-PS E246, (c) 
theoretical background (FSI),
(d) standard CKM model, (e) left-right symmetric models,
(f) multi-Higgs models with NFC, (g) multi-Higgs models without NFC,
(h) SUSY with R-parity, (i) SUSY without R-parity, and (j) leptoquark 
models.\\
\vspace{.5cm}
 
\renewcommand{\arraystretch}{2}
\noindent\begin{tabular*}{16.5cm}{@{\extracolsep{\fill}}ccccccccccccccr}
\hline \hline
&a & b&c& d&e& f&g &h&i&j\\
\hline
(1)&$10^{-2}$ & $5\cdot 10^{-4}$ &$ 10^{-6}$ & $0$ &0& $10^{-2}$&
$10^{-2}$ & $10^{-2}$ & $10^{-2}$ & $10^{-2}$\\ \hline
(2)& & $10^{-3}$ &$10^{-3}$& 0&$7\cdot 10^{-3}$& $5\cdot 10^{-3}$&$ 5\cdot
10^{-2}$ &$2\cdot 10^{-2}$&$1\cdot 10^{-2}$&$5\cdot 10^{-3}$\\ 
\hline \hline \end{tabular*}

\newpage       
\begin{figure}[h]
\includegraphics{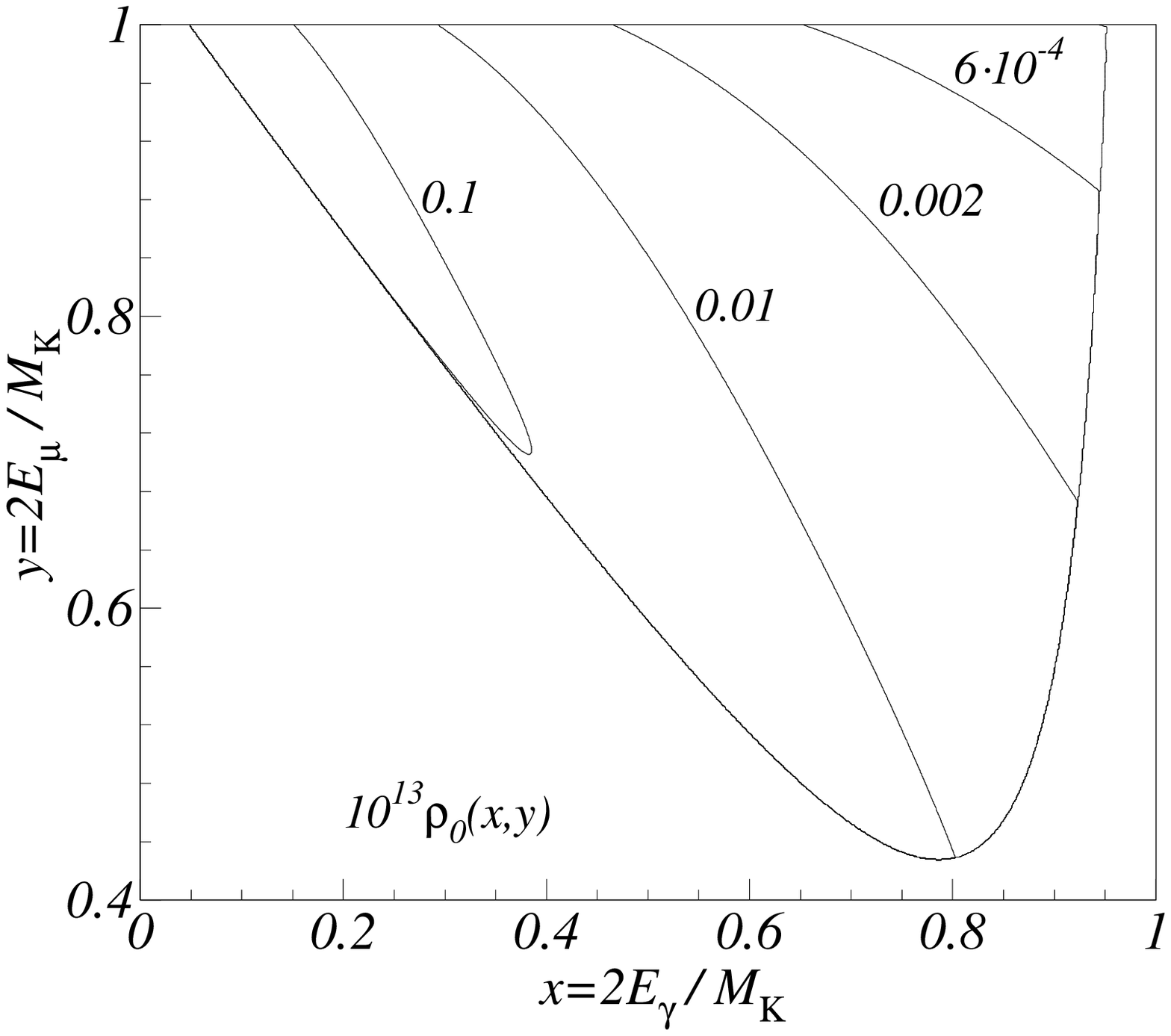}
\end{figure}
\vskip 9.1cm
\bc
Fig. 1.
Dalitz plot of $\rho_0(x,y)$.
\ec
\begin{figure}[h]
\includegraphics{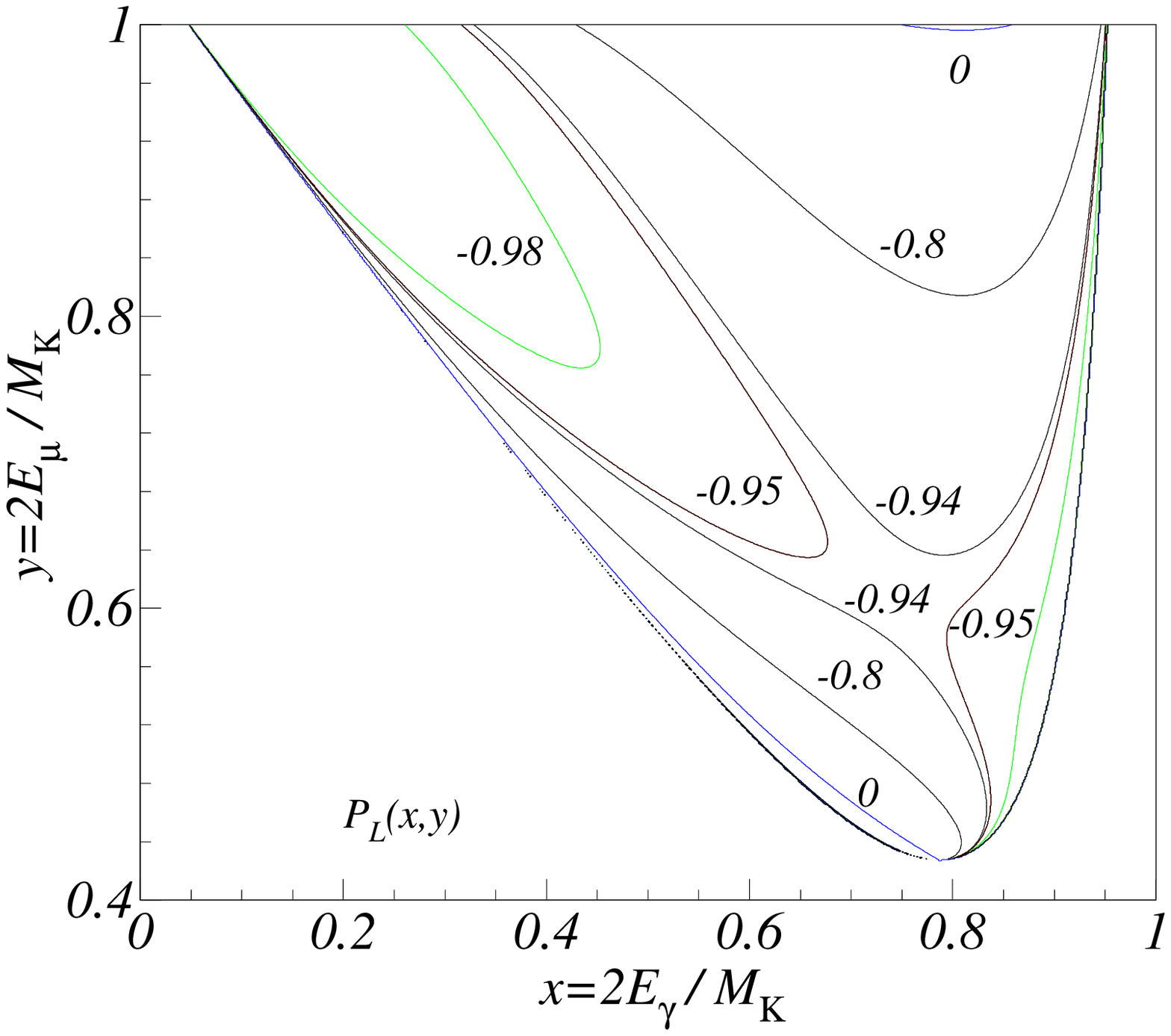}
\end{figure}
\vskip 10.5cm
\bc
Fig. 2.
Dalitz plot of $P_L(x,y)$.
\ec

\newpage       
\begin{figure}[h]
\includegraphics{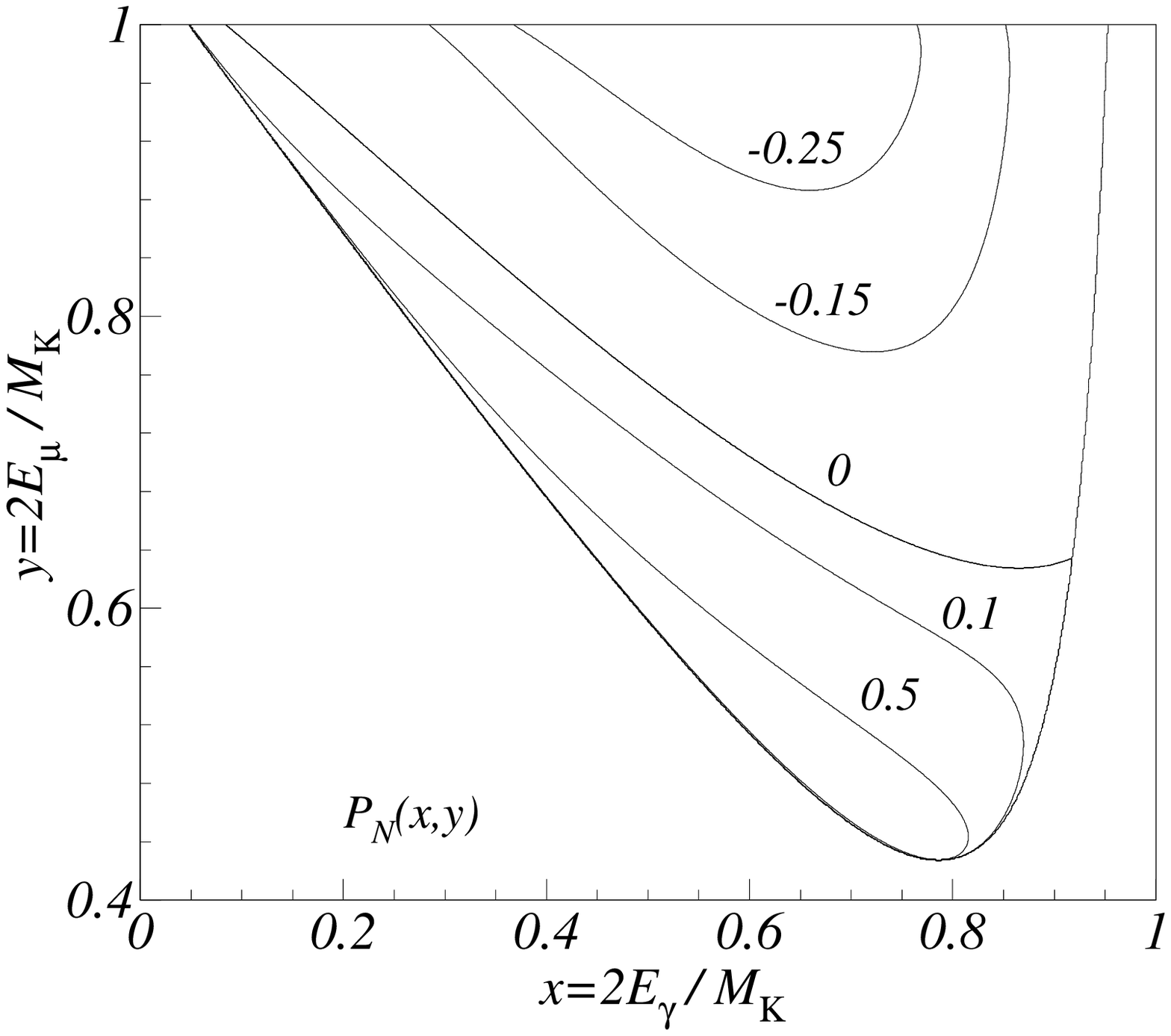}
\end{figure}
\vskip 9.1cm
\bc
Fig. 3.
Dalitz plot of $P_N(x,y)$.
\ec
\begin{figure}[h]
\includegraphics{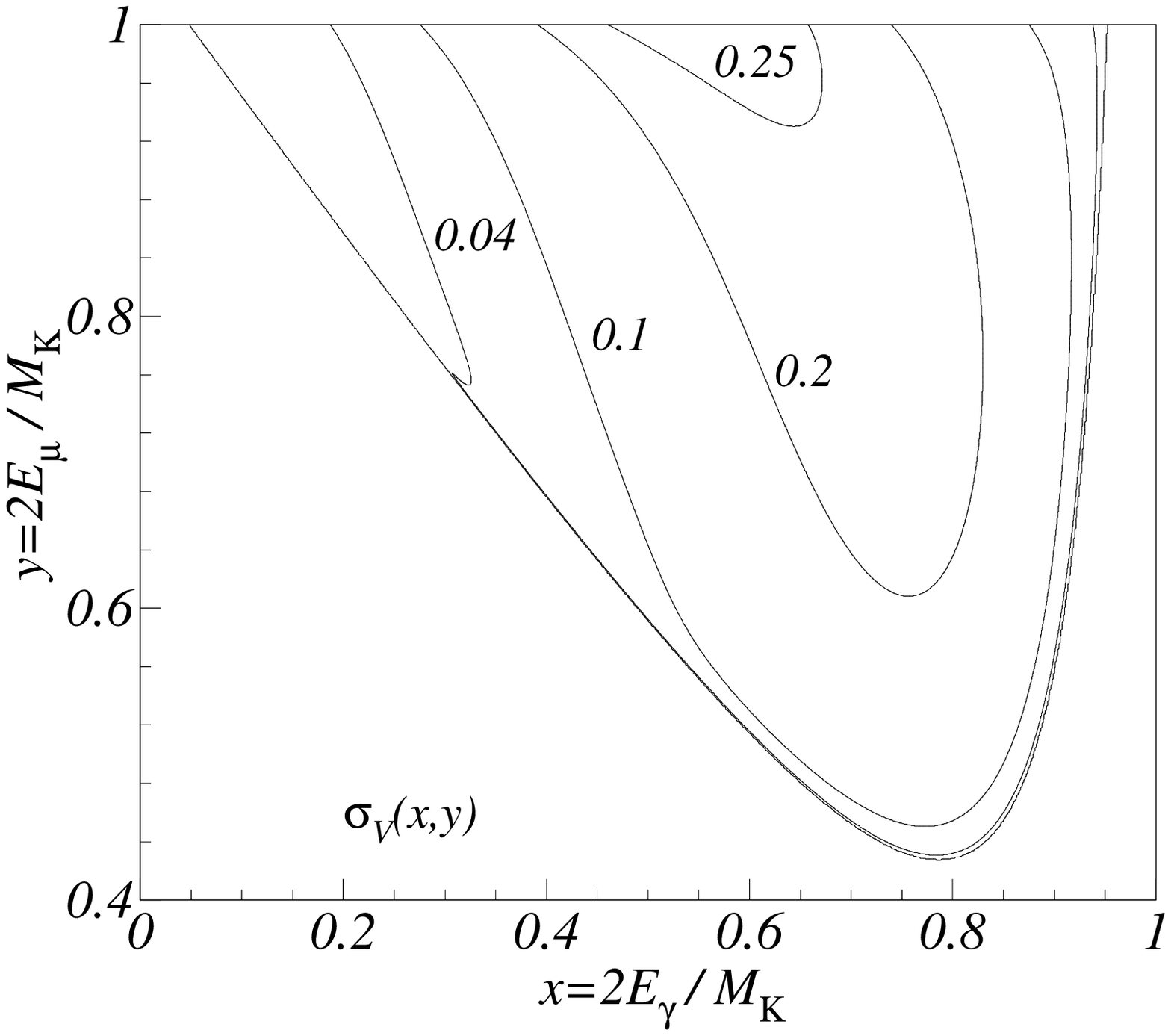}
\end{figure}
\vskip 10.5cm
\bc
Fig. 4.
Dalitz plot of $\sigma_V(x,y)$.
\ec

\newpage       
\begin{figure}[h]
\includegraphics{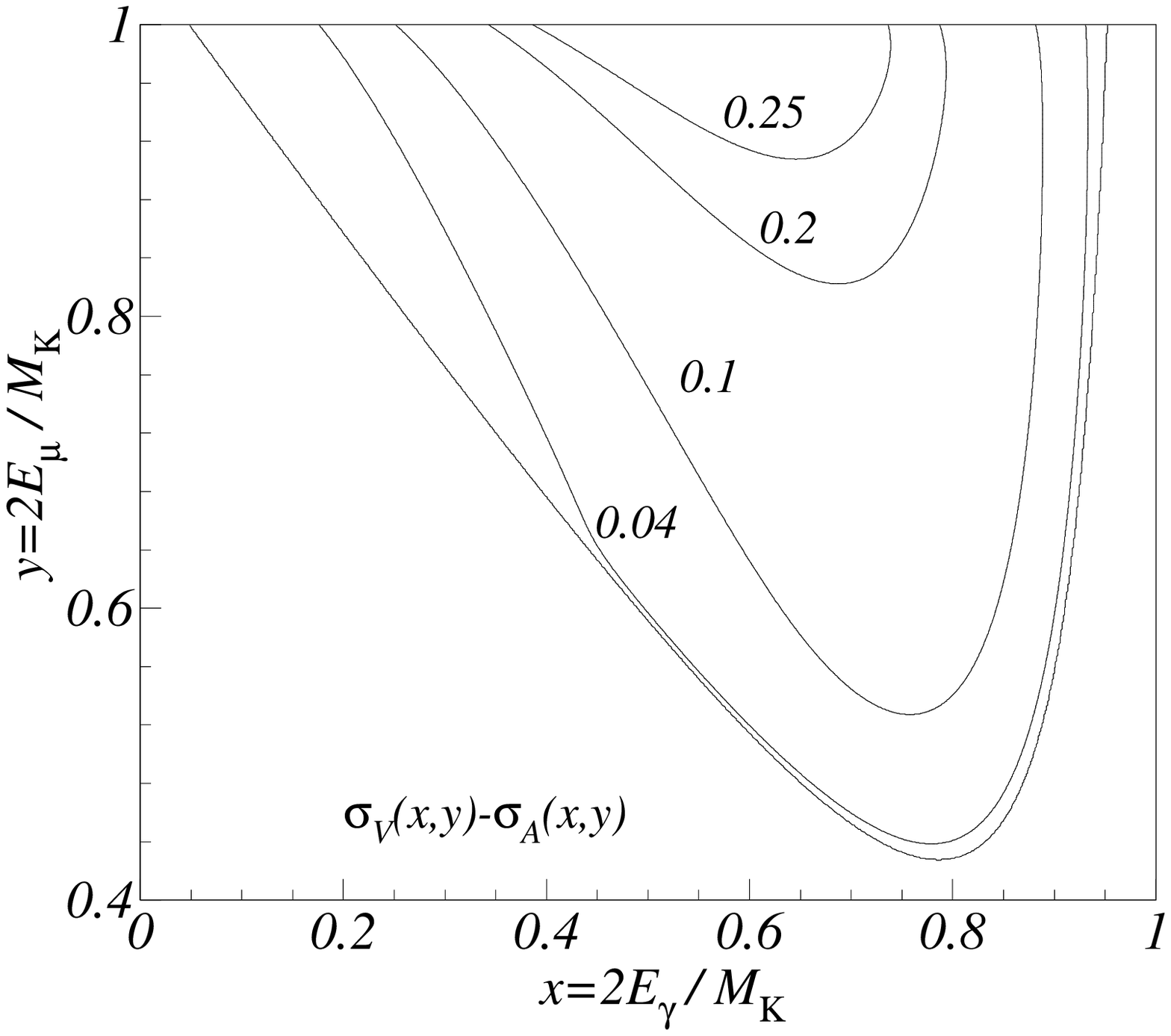}
\end{figure}
\vskip 9.1cm
\bc
Fig. 5.
Dalitz plot of $\sigma_V(x,y)-\sigma_A(x,y)$.
\ec
\begin{figure}[h]
\includegraphics{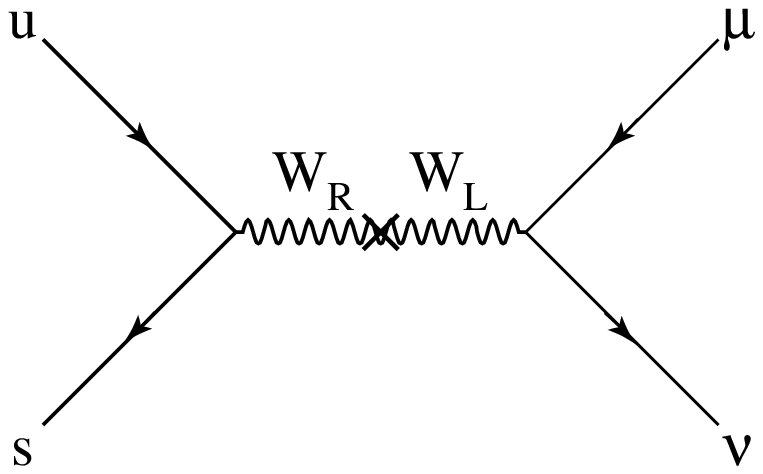}
\end{figure}
\vskip 9.5cm
\bc
Fig. 6.
Tree diagram to $\bar{s}u\to\mu\bar{\nu}$ 
induced by the left-right gauge boson mixing.
\ec

\newpage       
\begin{figure}[h]
\includegraphics{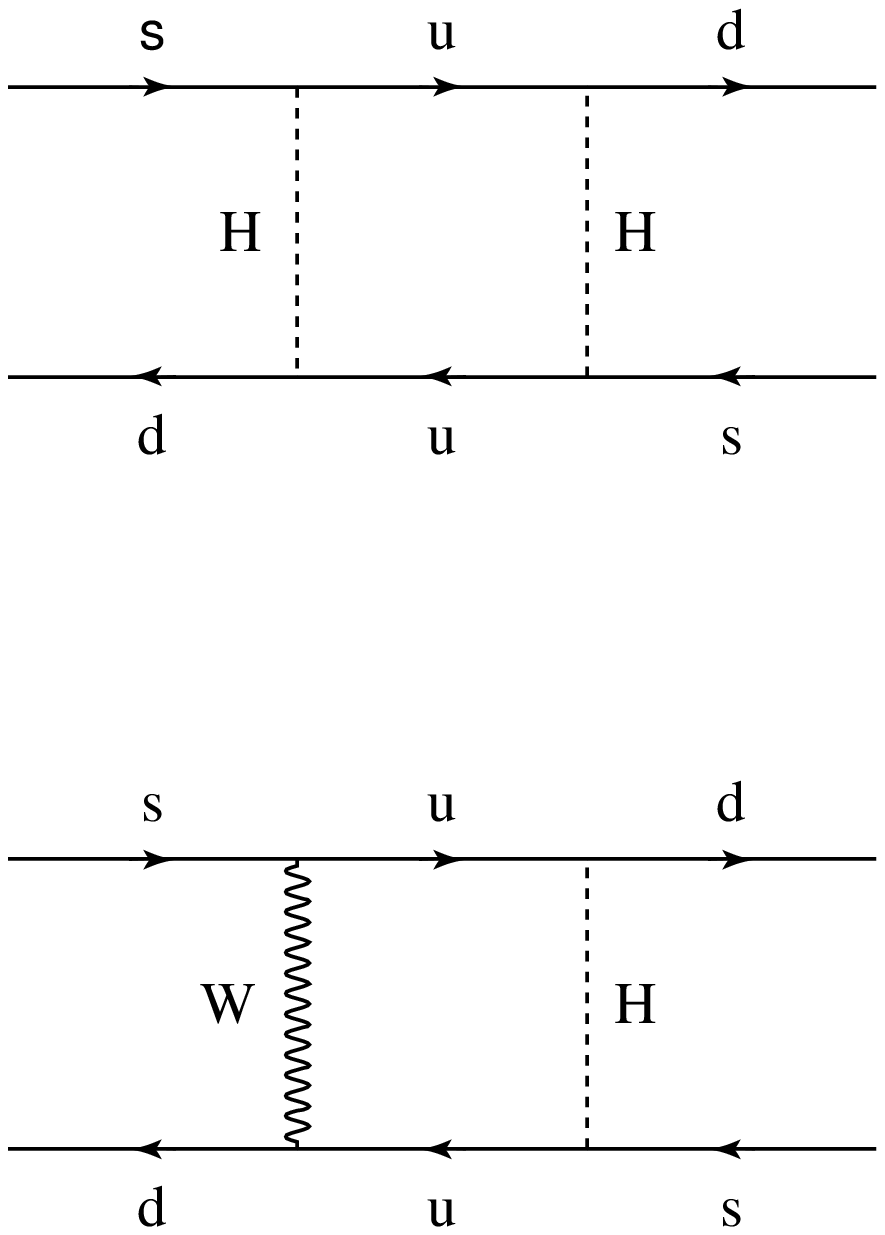}
\end{figure}
\vskip 9.1cm
\bc
Fig. 7.
Box diagrams to $K^0-\bar{K}^0$  mixing with the $u$-quark as the internal 
fermion. 
\ec
\begin{figure}[h]
\includegraphics{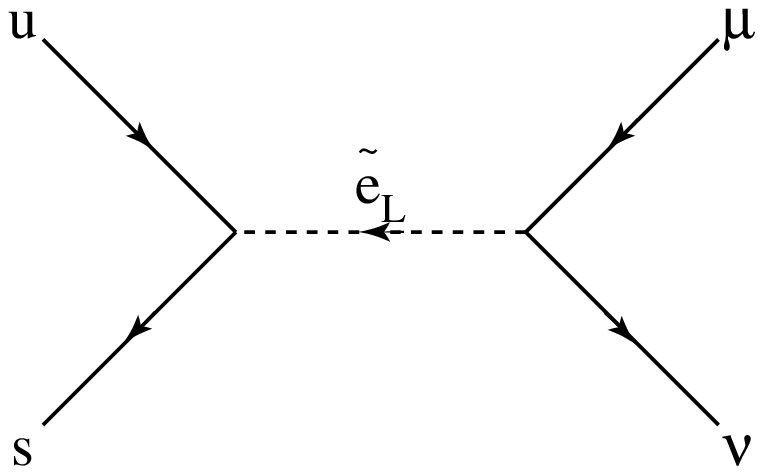}
\end{figure}
\vskip 9.5cm
\bc
Fig. 8.
Tree diagram to $\bar{s}u\to\mu\bar{\nu}$ with the slepton as 
the intermediate state.
\ec

\ed